\documentclass{article}

\usepackage{arxiv}
\usepackage[utf8]{inputenc}
\usepackage[T1]{fontenc}
\usepackage{amsmath}
\usepackage{amssymb}
\usepackage{graphicx}
\usepackage{booktabs}
\usepackage{tabularx}
\usepackage{microtype}
\usepackage[hidelinks]{hyperref}
\usepackage[numbers,sort&compress]{natbib}
\usepackage{url}

\title{A frozen rate operator from the complete larval connectome: degree and weight govern the gross response, exact wiring governs input routing and mushroom-body modes}
\renewcommand{\shorttitle}{A frozen rate operator from the larval connectome}
\renewcommand{\headeright}{Preprint}
\renewcommand{\undertitle}{Preprint}
\date{}

\author{%
  \href{https://orcid.org/0009-0007-1500-9779}{\includegraphics[scale=0.06]{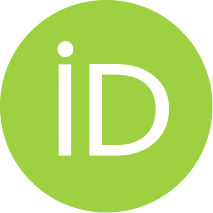}\hspace{1mm}Stavros Therianos} \\
  Independent Researcher \\
  \texttt{stherianos63@gmail.com} \\
}

\begin{document}
\maketitle

\begin{abstract}
Connectome-constrained models now reproduce neural activity in several systems, yet each inherits a circuit's degree and weight statistics along with its exact wiring, leaving open which dynamical properties the wiring fixes beyond those statistics. We separate the two by running the complete larval \textit{Drosophila} connectome, 2'825 neurons in its strongly connected core, as a frozen leaky-tanh rate operator with no single-neuron parameter fitted, and comparing it against a degree-and-weight-matched rewiring ensemble and a battery of further controls. Every property is the operator's, attributable to the wiring and weights alone, and is not a measurement of larval activity. The gross dynamical signature is largely fixed by coarse statistics: the operator is non-normal and near-linear, and its gain and effective dimensionality fall within a few percent of the ensemble. Two spatially resolved properties break this pattern. First, under sparse afferent drive, the operator confines activity to a fifth of the core, against two thirds for the ensemble and all of it for a random graph, a confinement that survives a cell-class-preserving null. Second, the mushroom body concentrates the leading driving modes far beyond the ensemble, surviving size-matched, singular-subspace, and family-wise controls, while its leverage elsewhere stays diffuse. Both depend on the exact placement of synapses. Taken together, we show here that a connectome's gross operator behavior is largely a property of its degree and weight statistics, while the routing of input and the identity of the dominant driving modes are written into its exact wiring. Beyond this connectome, we propose the decomposition as a general tool: it measures whether the exact wiring of any structured network carries information beyond its degree and weight statistics, a question that any use of biological connectivity as an architectural prior eventually faces.
\end{abstract}

\section*{1. Introduction}

Electron-microscopy reconstruction now yields the complete synaptic wiring diagram, the connectome, of an entire small brain. Complete reconstructions span the nematode \textit{C. elegans}, the larval and adult \textit{Drosophila} and larval zebrafish, and dense electron-microscopy volumes of mouse and human cortex are being assembled toward the same goal at mammalian scale \citep{winding2023connectome,cook2019celegans,dorkenwald2024flywire,hildebrand2017zebrafish,microns2025cortex,shapsoncoe2024human}. For the first time this makes it possible to ask, of a real brain rather than a model of one, which of its dynamical properties follow from the precise pattern of its connections and which follow from coarser statistical features of the wiring.

The question is more intricate than it first appears, for two reasons. First, connectivity constrains dynamics only loosely: networks with identical wiring but different cellular properties can produce different activity, so function cannot be read off a connectome alone \citep{beiran2025prediction}. Second, the choice of comparison decides what a result can mean. Showing that a connectome differs from an unstructured random network establishes that it is non-random, but it cannot tell whether a property depends on the exact placement of synapses or only on coarse features such as each neuron's number of partners and the distribution of its connection strengths. Separating the two requires a null model that preserves those coarse features and scrambles only the placement \citep{maslov2002specificity}.

The larval \textit{Drosophila} brain suits this question. It is the first complete insect connectome: 3'013 neurons joined by 111'243 directed connections, of which a strongly connected core of 2'825 neurons carries 109'438, in an animal whose behavior includes associative learning and navigation \citep{winding2023connectome}. It is small enough to analyze whole, yet it contains anatomically identified circuits with known functions: the mushroom body, the insect learning center; the lateral horn, a second-order olfactory center; the central complex, the navigation system; together with descending, sensory and convergence populations \citep{eichler2017mbconnectome,berck2016olfactory,eschbach2020recurrent,lungu2025larvalcx}. The same wiring has already been used as a fixed computational substrate, as a reservoir for time-series prediction \citep{suarez2024conn2res,costi2025reservoir} and, for the larval connectome, as a recurrent network trained only at its input and output \citep{yu2025bpu}. We instead characterize the operator itself, with no readout and no training: a frozen rate model in which no single-neuron parameter is tuned, so that every measured property is attributable to the wiring and its connection strengths. We report structural-dynamical properties of that operator, and make no physiological claim about activity in the living larva.

To draw the line between coarse statistics and exact placement, we read the operator against a battery of null models, each matched to the level of the claim it tests (Table 1). An unstructured Gaussian control, matched only in size and edge density, detects gross departures from randomness. A degree-and-weight-matched ensemble, which preserves every neuron's in-degree, out-degree and total outgoing weight together with the global multiset of connection strengths while scrambling where the connections go, tests whether a property needs the exact placement of synapses or only their coarse statistics \citep{maslov2002specificity}. For the claim that a particular circuit carries a property, two further controls ask whether the circuit is more extreme than arbitrary sets of the same size, and whether the effect is specific to the operator's eigenmodes rather than its high-gain directions. A property the matched ensemble reproduces is one the coarse statistics already fix. A property on which the connectome falls outside the ensemble is one its exact wiring fixes (Fig. 1).

\begin{figure}[tbp]
\centering
\includegraphics[width=\linewidth]{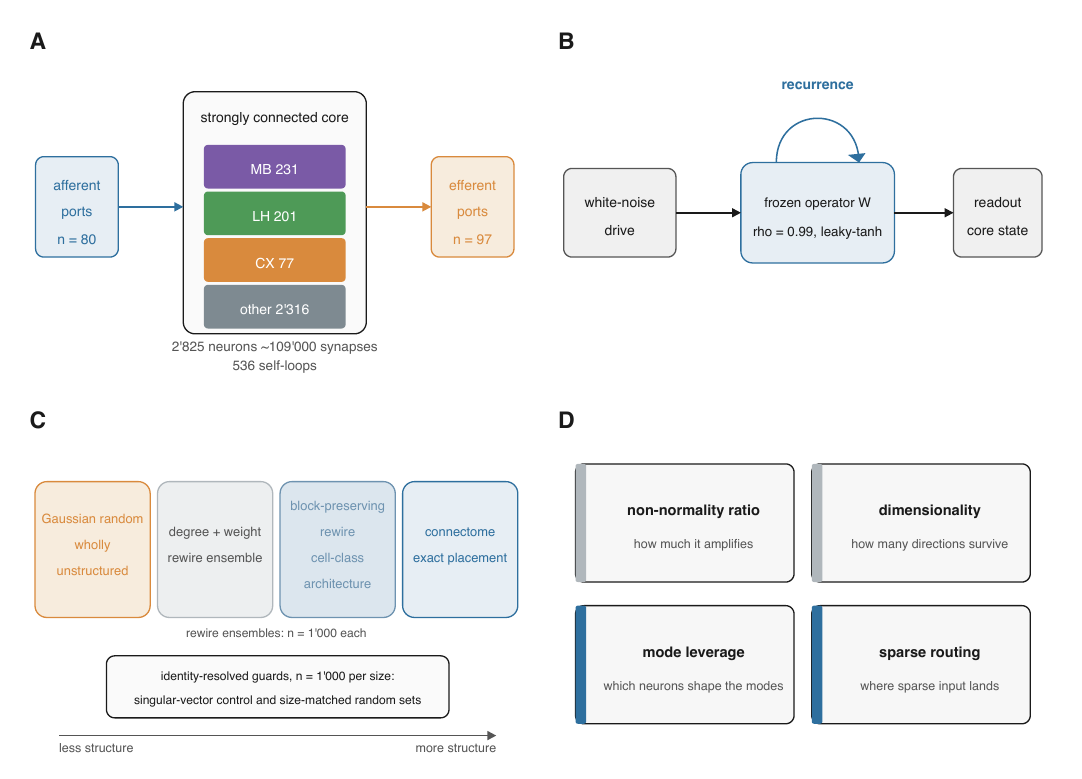}
\caption{\textbf{A complete connectome run as a fixed dynamical operator, read against a hierarchy of matched null networks.} \textbf{(A)} The recurrent core, 2'825 neurons joined by roughly 109'000 connections with 80 afferent and 97 efferent ports and 536 self-loops, containing anatomically identified circuits: the mushroom body (231 neurons), the lateral horn (201), the central complex (77) and the remaining population. \textbf{(B)} The frozen-operator assay. White-noise drive enters through the ports and propagates through the fixed weight matrix (spectral radius 0.99, leaky-tanh update) with no tuned single-neuron parameter, and the core state is read out. \textbf{(C)} The null-model ladder, from least to most wiring-preserving: an unstructured Gaussian control; the degree-and-weight-matched ensemble (n = 1'000); the block-preserving rewire that additionally fixes the cell-class architecture (n = 1'000); and the connectome. Identity-resolved claims add a size-matched random-set control and a singular-subspace guard. \textbf{(D)} The four properties probed: operator gain, dimensionality, mode leverage and sparse-input routing.}
\label{fig:1}
\end{figure}

Read this way, the operator separates into two kinds of property. First, its gross dynamical signature, how strongly and how richly it responds, is mostly degree-and-weight-governed: at a common spectral radius the matched ensemble reproduces it. Second, what the exact wiring fixes is where activity goes. Under sparse input the connectome confines activity to a fifth of the core and routes it into the olfactory pathway, while matched rewiring floods the brain. On the driving side of the dynamics, the mushroom body concentrates the operator's leading modes beyond every matched control. A candidate localization in the lateral horn did not survive the size-matched control and is retracted, an instance of these controls working. We report these as properties of a structural-dynamical instrument: it omits action potentials, single-neuron time constants, synaptic and receptor kinetics, neuromodulatory and behavioral state, and gap junctions, and makes no claim about activity in the living animal.

\begin{table}[tbp]
\centering
\caption{\textbf{The control battery.} Each control is matched to the level of the claim it licenses, from non-randomness through exact synaptic placement to circuit-level specificity.}
\label{tab:controls}
\begin{tabularx}{\linewidth}{@{}p{0.17\linewidth}XX@{}}
\toprule
\textbf{Control} & \textbf{What it tests against} & \textbf{What a pass establishes} \\
\midrule
Unstructured Gaussian & A random matrix matched only in size, edge density and spectral radius & The property departs from what an unstructured network of the same size produces, establishing that it is non-random \\
Degree-and-weight-matched ensemble & 1'000 rewirings preserving every neuron's in-degree, out-degree and out-strength and the global weight multiset, scrambling placement & The property requires the exact placement of synapses, beyond the coarse degree and weight statistics \\
Block-preserving (cell-class) null & Rewirings that additionally fix the class-to-class edge-count matrix, scrambling placement only within cell-class blocks & The property requires finer, within-class placement, beyond cell-class architecture \\
Random size-matched sets & 1'000 arbitrary neuron sets of the named circuit's size & The property is specific to the named circuit rather than to any set of that size \\
Singular-subspace guard & The operator's leading singular (high-gain) subspace & The effect is specific to the operator's eigenmodes rather than a feature of its high-gain directions \\
Westfall-Young family-wise control & A max-statistic permutation across the full 105-entry panel & The effect survives correction for scanning every circuit, subspace and mode count \\
\bottomrule
\end{tabularx}
\end{table}

\section*{2. Results}

\subsection*{2.1 Global operator behavior is mostly governed by degree-and-weight statistics}

We first asked whether the recurrent core's gross dynamical behavior comes from the precise placement of its synapses or from coarser features of the wiring. The distinction sets how much a connectome-constrained model owes to the exact circuit and how much to statistics any wiring with the same degree and weight profile would share. Run as a frozen operator, the core is low-gain and only weakly non-normal: it amplifies activity modestly, and amplifies somewhat more in a single step than its long-run gain would suggest. We summarize this by the ratio of the largest single-step amplification to the asymptotic gain, $\sigma_1/\rho$, which is 1.325 for the connectome (spectral radius $\rho = 296.6$, largest singular value $\sigma_1 = 393.1$, Frobenius norm 1'825, before the assay's spectral-radius normalization). The degree-and-weight-matched ensemble, which preserves each neuron's in-degree, out-degree and outgoing weight and the global weight multiset while scrambling placement, reproduces this ratio closely (1.309 $\pm$ 0.007); the connectome sits high within the ensemble, at rank 980 of 1'001 ($z = +2.2$, with 2.1\% of rewires above it). An unstructured Gaussian operator is far higher (2.079) (Fig. 2a,b). The operator's modest gain is therefore governed by the wiring statistics: scrambling the placement while preserving those statistics leaves it essentially unchanged \citep{hornjohnson2012matrix}.

\begin{figure}[tbp]
\centering
\includegraphics[width=\linewidth]{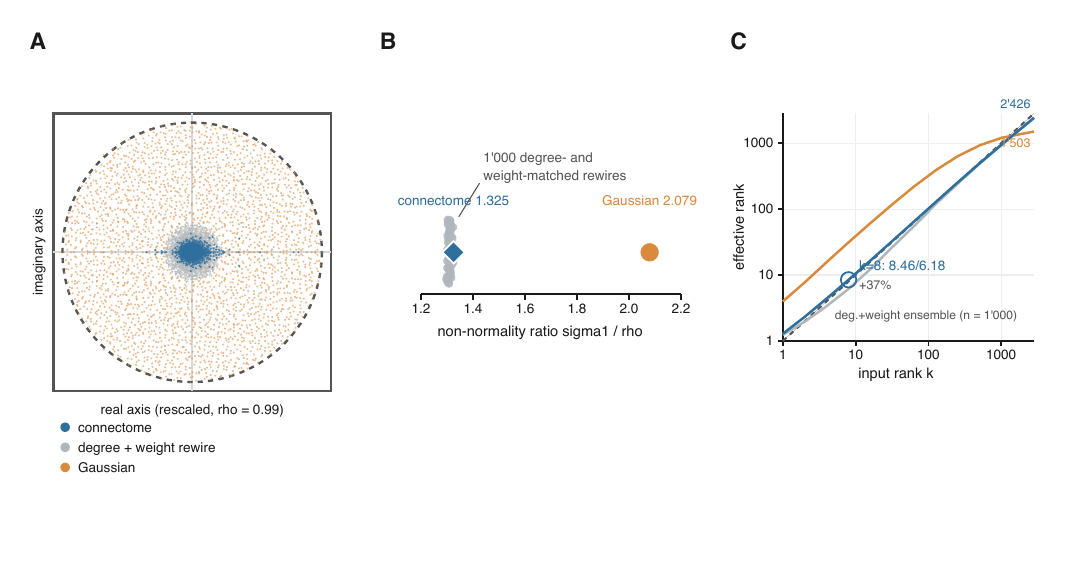}
\caption{\textbf{Global operator behavior is mostly statistical.} \textbf{(A)} Eigenvalue spectrum of the rescaled operator: the connectome's eigenvalues cluster near the origin (blue), whereas an unstructured Gaussian operator fills the spectral disk (orange). \textbf{(B)} Non-normality ratio, the ratio of the largest single-step amplification to the asymptotic gain ($\sigma_1/\rho$). The connectome (1.325) sits within the degree-and-weight-matched ensemble (1.309 $\pm$ 0.007, n = 1'000, rank 980 of 1'001), far below the Gaussian control (2.079). \textbf{(C)} Effective rank versus input rank $k$. The connectome (blue) preserves dimensionality, reaching 2'426 at full forcing, and the degree-and-weight ensemble (grey) tracks just below it (6.18 against the connectome's 8.46 at $k = 8$, a 37\% increment), whereas the Gaussian control (orange) expands then saturates at 1'503. $\sigma_1$, largest singular value; $\rho$, spectral radius.}
\label{fig:2}
\end{figure}

We state this as an empirical reproduction rather than a proof of sufficiency. The ensemble preserves each neuron's in-degree, out-degree and out-strength, its column sum because weights travel with their source, and the global weight multiset, but not each neuron's in-strength, its row sum. Because out-strength fixes each column sum, the matrix 1-norm is preserved across the ensemble by construction, so this reproduction of the gain ratio is in part definitional; the dimensionality and near-linearity reproductions, which the 1-norm does not fix, are not. This null is a strong test of placement beyond degree and outgoing weight, and a weaker test of incoming weighted mass, which is why the routing analysis below adds a row-strength diagnostic.

Dimensionality follows the same pattern. Effective rank, the number of independent activity directions an operator preserves, is far higher for the connectome (2'426 of a possible 2'825 dimensions) than for the Gaussian control (1'503), and here too the matched ensemble, not the random control, is the connectome's near neighbor. The precise wiring leaves only a modest residual. Driven at low input rank ($k = 8$), the connectome preserves slightly more dimensions than its rewires (effective rank 8.46 against 6.18), an excess of roughly 37\% that shrinks to under 1\% once the full input space is engaged. The operator is also marginally more linear than its rewires: its high-rank nonlinearity is 0.0179, below every one of the 1'000 matched rewires (ensemble 0.0323 $\pm$ 0.0002) (Fig. 2c). These wiring-specific differences are real but small. Because the matched ensemble's spread is so narrow, a bare rank would overstate them, so we report effect sizes instead. The per-statistic ensemble distributions behind every reported effect size are given in Supplementary Table S1.

The three measures agree. At a common spectral radius, the connectome's gain, its dimensionality and its degree of nonlinearity are mostly set by its degree-and-weight statistics, leaving a wiring-specific increment of at most roughly 37\% at low and middle dimensions. The exact wiring's decisive effects lie elsewhere, in where activity is routed, which we take up next. The value of the matched null is that it makes this separation quantitative, fixing how much of the gross response the coarse statistics already account for.

\subsection*{2.2 Module maps recover known circuits, with definition-dependent boundaries}

We next asked whether the connectome's communities correspond to the brain's known circuits, and how cleanly each circuit can be drawn. Because the later analyses ask whether particular circuits carry particular dynamical properties, we first define the circuits and state how sharply each is bounded. We found that the mushroom body, the insect learning center, is the sharpest. Under dataset cell-class labels it comprises 153 Kenyon cells, the parallel-fiber neurons that store associations, together with 48 output neurons, 28 modulatory input neurons and the two anterior paired lateral (APL) neurons, 231 in total \citep{eichler2017mbconnectome}. The module spans both hemispheres, so the two APLs are the bilateral pair, one per mushroom body, each the single GABAergic interneuron that provides feedback inhibition to its Kenyon cells \citep{masuda2014apl,mancini2023apl}.

The lateral horn and the central complex are bounded less sharply, and the boundary depends on how the circuit is defined, by connectivity clustering or by expert anatomical annotation. For the lateral horn, the second-order olfactory center, the clustering-defined core contains 201 neurons, the annotation-defined set 445, and their intersection 164 \citep{berck2016olfactory,eschbach2020recurrent}. For the central complex, the navigation system, the clustering-defined set contains 77 neurons, the annotation-defined set 34, and their overlap 10 \citep{lungu2025larvalcx}. The two lateral-horn sets are defined within the strongly connected core, while the 34-neuron central-complex annotation set is drawn from the full reconstructed connectome; all 34 of its neurons nonetheless lie in the core. Every later claim about these circuits therefore states which boundary it uses, and we treat no central-complex result as a boundary-free statement about a single unambiguous set. The modular block structure these definitions partition is visible in the connectome's community-ordered connectivity matrix (Fig. 3a), and the full set-size comparison is given in Supplementary Table S5, with the module-set boundaries shown in Supplementary Figure S1. Connectivity-defined communities in this connectome align with cell type and function while their boundaries shift with the detection method \citep{betzel2024hierarchical}.

\begin{figure}[tbp]
\centering
\includegraphics[width=\linewidth]{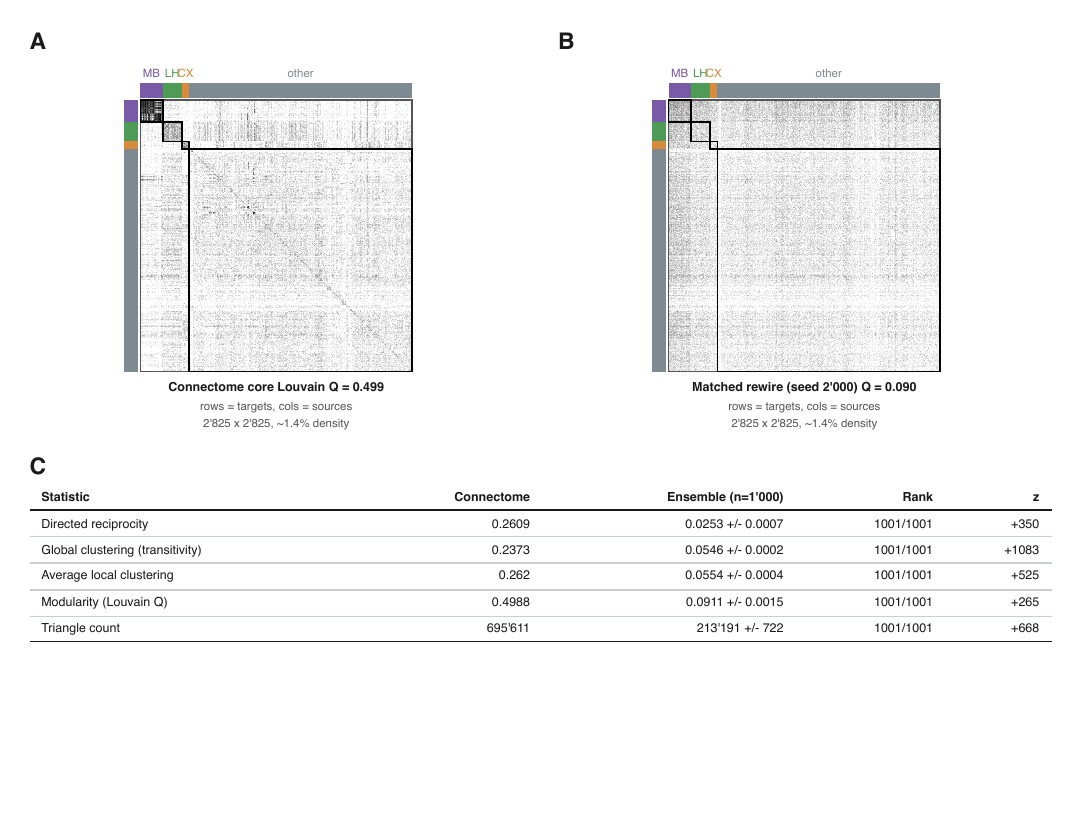}
\caption{\textbf{The matched ensemble destroys the connectome's higher-order structure while preserving its degree and weight.} \textbf{(A)} Community-ordered connectivity matrix of the connectome (Louvain modularity Q = 0.499; 2'825 x 2'825, about 1.4\% density), with the mushroom-body, lateral-horn and central-complex blocks outlined. \textbf{(B)} The same for a degree-and-weight-matched rewire (Q = 0.090); the block structure is gone. \textbf{(C)} Five higher-order statistics, connectome against the n = 1'000 ensemble: directed reciprocity, global and average-local clustering, modularity and triangle count each stand far above the entire ensemble (rank 1'001 of 1'001). Q, Louvain modularity.}
\label{fig:3}
\end{figure}

As a structural check, the module maps recover the expected anatomy of olfactory-to-convergence flow: both the mushroom-body and lateral-horn output populations route prominently onto convergence neurons, the cells that pool signals from several pathways (Methods 4.7). Outside the three modules, the remaining 2'316 core neurons fall, by dataset class, into a majority of typed but otherwise unclassified neurons (1'137, un-typed rather than unimaged), then descending (349), sensory (338), convergence (184), feedback (142) and olfactory projection and local neurons (114), with 52 further typed. These classes provide the cell-class vocabulary for the routing analysis below. The connectome's communities do recover the known circuits, but only the mushroom body has a hard edge. The lateral horn and the central complex are drawn differently depending on whether one follows the wiring or the anatomist, which is why every result that follows names the boundary it stands on.

\subsection*{2.3 The mushroom body concentrates the driving modes; mode leverage is otherwise diffuse}

We next asked whether the precise wiring affords any single circuit a disproportionate role in the operator's dominant dynamical modes, beyond what its degree and weight statistics would predict, or whether that influence is spread across the brain. A dynamical mode has two sides, and we separate them throughout. The driving side (the left eigenvectors, equivalently the leading subspace of the adjoint operator) is the set of neurons that weigh most heavily in a mode. The driven side (the right eigenvectors) is the set the mode most loads onto (Fig. 4a). Because the operator is non-normal, the two sides differ. Concentration on a circuit's left-eigenvector subspace is a spectral property of the wiring. It does not by itself establish causal driving or input-output influence, which would require a controllability analysis we do not attempt. We use the term driving in this spectral, adjoint-subspace sense throughout. We tested seven pre-specified neuron sets: five anatomically defined circuits, the mushroom body, the lateral horn, the central complex, the convergence neurons and the descending output neurons; the mushroom body's recurrent output network, formed by its output and modulatory neurons; and the population the connectome activates under sparse input drive (defined in the routing analysis below). For each set we measured its leverage, the share of the leading modes' energy that concentrates on it, across mode counts from 1 to 16, and required a candidate to pass two controls. The first control asks whether the circuit is more extreme than arbitrary sets of the same size. We compare it against 1'000 random size-matched sets and keep an entry only when the connectome exceeds their 95th percentile and the random sets are themselves extreme in fewer than 20\% of draws. The second control asks whether the concentration is specific to the operator's eigenmodes or merely a feature of its high-gain directions. We check that the leading singular subspace, the operator-norm subspace of maximal input-output gain, does not carry the same concentration. The two controls together separate a wiring-specific eigenmode role from one the operator's coarse statistics or its gain geometry would produce on their own.

\begin{figure}[tbp]
\centering
\includegraphics[width=\linewidth]{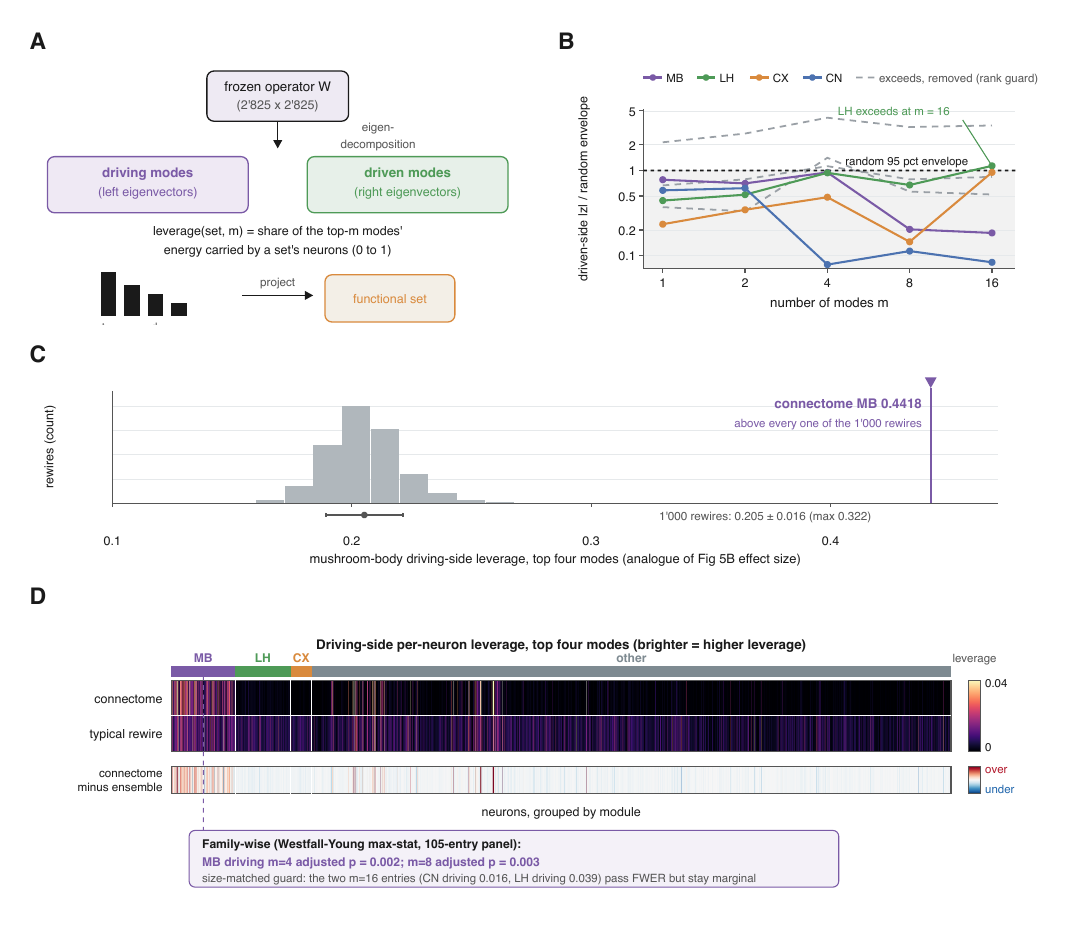}
\caption{\textbf{The mushroom body over-concentrates the driving modes; mode leverage is otherwise diffuse.} \textbf{(A)} Leverage is the share of a set of top-$m$ modes' energy carried by a neuron set, on the driving (left-eigenvector) and driven (right-eigenvector) subspaces. \textbf{(B)} Driven side. For each circuit the connectome's deviation lies at or below the 95th-percentile envelope of 1'000 random size-matched sets across mode counts $m$; the lateral horn exceeds it only at $m = 16$, where same-size random sets are themselves extreme in 36\% of draws, and three further sets are flagged generic by the singular-subspace guard, so no set clears both guards and the driven side is diffuse. \textbf{(C)} Driving side, quantitative. The mushroom-body driving-side leverage at four modes (0.4418) lies above every one of the 1'000 degree-and-weight rewires (mean 0.205, s.d. 0.016, maximum 0.322). \textbf{(D)} Driving-side per-neuron leverage of the top four modes for the connectome and a typical rewire, and their difference (red, over-concentrated; blue, under). The connectome concentrates leverage on the mushroom body. A family-wise permutation control (Westfall-Young max-statistic over the 105-entry panel) gives the mushroom-body entries adjusted p = 0.002 ($m = 4$) and 0.003 ($m = 8$); two further entries pass at $m = 16$ but remain marginal on the size-matched guard. $m$, number of modes; s.d., standard deviation.}
\label{fig:4}
\end{figure}

The result is diffuse across most of the circuits, with a small structured set of departures on the driving side (Fig. 4). Five entries meet the pre-specified guard and are listed in Supplementary Table S4. Three are comfortably inside it and carry the result. The strongest is the mushroom body, the learning center, on the driving side. Its leading-mode concentration exceeds the entire 1'000-rewire ensemble at four modes (0.4418, against an ensemble mean of 0.2053 $\pm$ 0.0161 and maximum of 0.3222; $z = +14.7$, rank 1'001 of 1'001) and at eight modes (0.3163, against 0.1935 $\pm$ 0.0133, maximum 0.2474; $z = +9.2$) (Fig. 4c). Both survive the size-matched control with wide margin (random sets extreme in 4.7\% and 7.1\% of draws), and the concentration is not merely a gain-subspace effect: on the driving side it far exceeds what the singular subspace produces ($z = +14.7$ and $+9.2$ on the driving side, against $+2.9$ and $+1.5$ for the singular subspace), and the singular subspace is not itself at the extreme rank. Both reproduce exactly under a dense eigendecomposition, and the driving-side subspace is numerically well-conditioned despite the operator's non-normality (Methods 4.10). The operator's driving modes concentrate on the mushroom body above the matched ensemble under both guards. In other words, the learning center's exact wiring lifts it into the set of neurons that weigh most heavily in the circuit's leading modes. The convergence neurons give a second, weaker entry at eight modes, in the opposite direction (an under-concentration, $z = -5.2$), so the wiring concentrates driving influence in the mushroom body and away from the convergence pool. Two further entries, both at the highest mode count, sit at the edge of the size-matched threshold (random sets extreme in 19\% of draws, against the 20\% bound) and we do not lean on them. The central complex, whose boundary is itself definition-dependent, shows no certified entry, and we make no wiring-specificity claim for it.

The clearest negative comes from the lateral horn, on the driven side of the dynamics. An earlier analysis had flagged the lateral horn as a candidate on the driven side, where it appeared to take an unusually small share of the leading modes relative to the rewired ensemble. The size-matched control, at 1'000 sets, removes it. For driven-side mode counts 1 through 16, the lateral-horn deviation $|z| = [18.78, 6.84, 7.87, 6.88, 7.80]$ lies below the random size-201 95th-percentile envelope $[42.45, 13.16, 8.40, 10.19, 6.89]$ (Fig. 4b) at every count through eight. At sixteen modes it rises just above the envelope (7.80 against 6.89), but random sets of the same size are themselves extreme in 36\% of draws, far above the 20\% bound, so the control does not certify the lateral horn there either. We therefore retract the lateral-horn localization and fold it into the diffuse result. The retraction is on the driven side and the surviving mushroom-body result is on the driving side, so the two concern opposite sides of the operator and are not in tension.

Mode leverage is therefore mostly diffuse: across seven circuits and five mode counts, the exact wiring gives almost no circuit a dominant-mode role that its statistics do not already explain. The complete matrix of all seven sets, both eigenvector sides and the singular subspace across the five mode counts, 105 entries, is reported in Supplementary Table S3, so the certified entries are read against every test performed. Five entries pass the guard and three carry the result. Across that 105-entry family a Westfall-Young max-statistic permutation over the degree-and-weight ensemble controls the family-wise error: the mushroom-body driving-side entries survive at adjusted $p = 0.002$ ($m = 4$) and 0.003 ($m = 8$) (Fig. 4d), and although all five flagged entries fall below an adjusted threshold of 0.05, the two highest-mode-count entries remain marginal under the size-matched guard, so we lean only on the three inside both controls. The robust departures form a small structured set on the driving side: the mushroom body over-concentrates the leading adjoint modes beyond every matched control, and the convergence neurons under-concentrate them. Almost nowhere does the precise wiring single out a circuit beyond its raw statistics, and the one place it clearly does is the learning center, whose wiring concentrates the circuit's most influential modes onto itself.

\subsection*{2.4 Specific wiring confines sparse input into the olfactory pathway, beyond cell-class architecture and incoming strength}

We then asked whether the exact placement of synapses controls where input-driven activity travels, or whether the connectome's coarse wiring statistics already fix it. If placement matters, we further asked whether the grouping of neurons into cell classes does the work or a still finer pattern within those classes. The clearest wiring-specific result comes from routing rather than from the dominant modes. We drove the core through its afferent ports, an abstract input shell, and measured how widely activity spreads. Here activity means the modeled operator state variance each neuron carries under the drive, and a neuron counts as active when that variance exceeds the threshold defined in Methods 4.9. The larval olfactory receptor neurons lie inside the modeled core rather than on this afferent shell, so the drive enters the olfactory pathway downstream of the receptor layer. The routing below therefore refers to which modeled cell classes carry the propagated state, and not to odor-evoked activity in the larva. The connectome confines the drive to a fifth of the core (561 of 2'825 neurons, an active fraction of 0.1986), while the degree-and-weight-matched ensemble lets the same drive spread to two-thirds (0.6682 $\pm$ 0.0078) and an unstructured Gaussian operator with no wiring structure activates essentially the whole core (1.0). The connectome is the most confined of all 1'001 graphs (rank 1, $z = -60.56$), a 3.4-fold deepening relative to its matched rewires (Fig. 5a,b).

\begin{figure}[tbp]
\centering
\includegraphics[width=\linewidth]{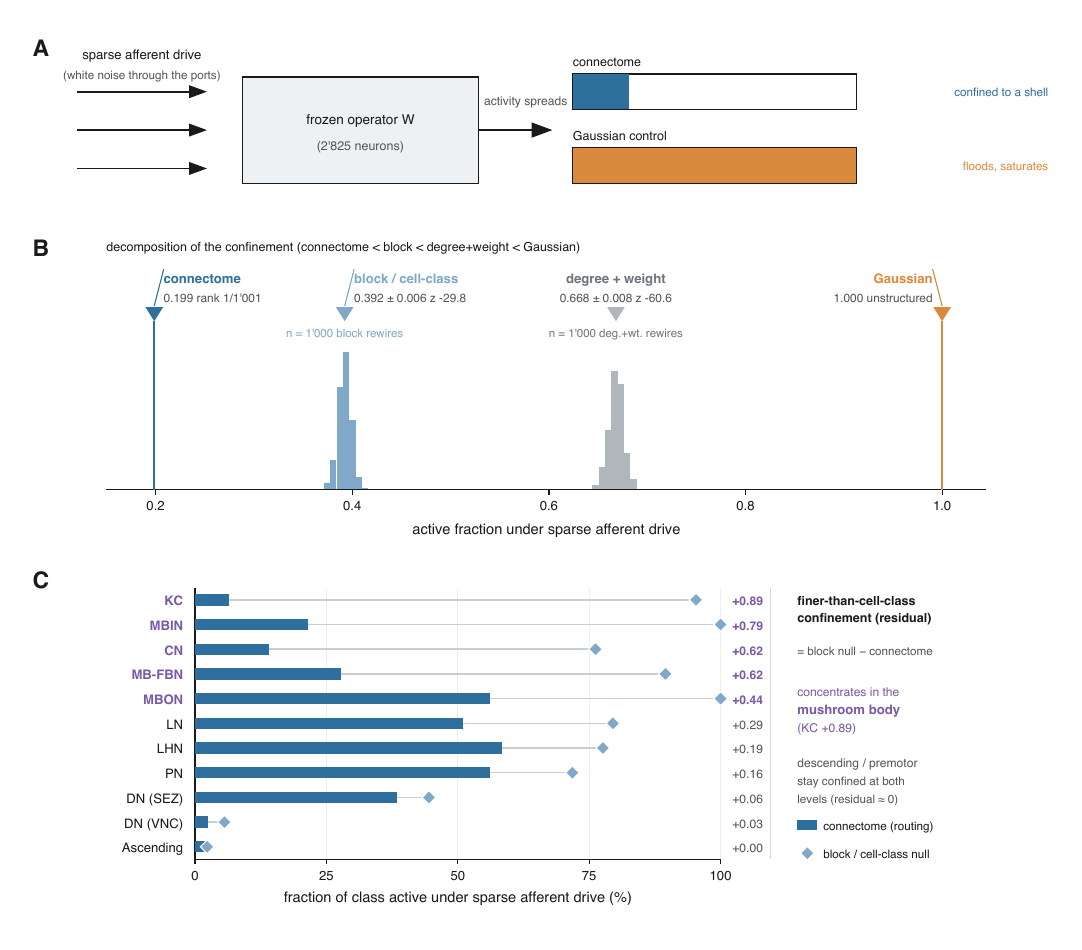}
\caption{\textbf{The exact wiring confines sparse input to the olfactory pathway, and the confinement separates into a cell-class and a finer component.} \textbf{(A)} Sparse white-noise drive enters the core through its afferent ports and propagates through the fixed operator; the fraction of the core that becomes active separates the connectome from its randomized controls. \textbf{(B)} Active fraction under sparse drive, as a decomposition ladder. The connectome activates a fifth of the core (0.199, rank 1 of 1'001, z = -60.6), the degree-and-weight ensemble two-thirds (0.668), and the Gaussian control the whole core (1.0); a block-preserving null that additionally fixes the cell-class architecture lies between (0.392, rank 1 of 1'001 against it, z = -29.8), so the confinement decomposes. The ensemble reproduces the connectome's non-normality ratio (1.31 against 1.325), so the deeper confinement reflects wiring rather than gain. \textbf{(C)} Finer-than-cell-class confinement by cell class (block-null active fraction minus connectome). The residual concentrates in the mushroom-body classes (Kenyon cells +0.89; mushroom-body input, feedback and output neurons and the convergence neurons +0.79 to +0.44) and is near zero for the descending and premotor classes, which the cell-class architecture already confines. z, standardized deviation from the matched ensemble.}
\label{fig:5}
\end{figure}

The drive does not spread evenly. It concentrates in the olfactory pathway: projection and local neurons (55\% of them active), the lateral horn (52\%) and the convergence neurons (42\%) take most of it, while the mushroom body is largely spared (19\% active) and the central complex receives none (0\%). The reading is channeling rather than broadcast: even the most engaged class sits just above half, and the drive falls to a fifth in the mushroom body and to nothing in the central complex, so the wiring meters the drive everywhere it reaches and never saturates the regions it enters. The matched rewires flood all of these instead, activating 98\% of the mushroom body, 90\% of the lateral horn and 77\% of the central complex in a representative rewire (Fig. 5). The specific wiring thus selects which neurons receive the drive, channeling it into a functionally coherent olfactory shell, a selection its degree and weight statistics do not make.

Two facts establish that this is wiring-specific rather than a side effect of overall gain. First, the matched ensemble reproduces the connectome's non-normality ratio (1.31 against 1.325), so the deeper confinement cannot follow from how strongly the operator amplifies. Second, the same ensemble, which preserves every neuron's degree and out-strength and the entire weight multiset, destroys the connectome's higher-order structure: directed reciprocity (0.2609 against an ensemble 0.0253), global and local clustering (0.2373 against 0.0546; 0.2620 against 0.0554), modularity (0.4988 against 0.0911) and triangle count (695'611 against 213'191) all stand far above the entire ensemble, each at rank 1'001 of 1'001 (Fig. 3b,c). The connectome carries recurrent structure, loops, clustering and modules, that degree and weight alone do not produce, consistent with the nested recurrent loops and multilayer shortcuts mapped in the larval brain \citep{winding2023connectome}, and that structure is what routes the drive.

To localize the structural level of this confinement, we then held the cell-class architecture fixed and scrambled only the placement within it. Cell-class architecture here means the grouping of neurons into types together with the connection counts between those types. The block-preserving null preserves, against the connectome, every neuron's in-degree and out-degree, each neuron's out-strength, the full synaptic-weight multiset, and the 18-by-18 class-to-class edge-count matrix, while rewiring freely inside each cell-class block (edge displacement 0.835, near-complete within-block mixing). Keeping which neuron types wire to which therefore recovers part of the confinement on its own: under the same sparse drive the block-preserving ensemble activates 39.2\% $\pm$ 0.7\% of the core, between the connectome's 19.9\% and the degree-and-weight ensemble's 66.8\% ($n$ = 1'000 each; Supplementary Table S2). The two levels are comparable in size: cell-class architecture produces the reduction from 66.8\% to 39.2\% and the finer within-class placement the reduction from 39.2\% to 19.9\%, so the finer share of the confinement beyond degree and weight is about 41\% by active fraction and Gini and about 55\% by participation ratio (Fig.~6 partitions the total confinement instead, where this finer component is the 24\% residual). The connectome stays far more confined than the block-preserving null at every instance (rank 1 of 1'001, $z = -29.8$), across all eight activity thresholds from 0.02 to 0.50, across an excitation-inhibition sign cone (20 polarity draws, the connectome below both signed ensembles), and across a grid of spectral radius, leak and input amplitude (Supplementary Table S9). The confinement therefore decomposes: cell-class architecture confines the drive partway, beyond degree and weight, and the finer within-class wiring confines it further still.

The finer component falls mostly on one circuit. Resolving the block-null active fractions by cell class, the residual between the block-preserving ensemble and the connectome concentrates in the mushroom body (Fig. 5c). Cell-class architecture alone would activate the Kenyon cells to 95\%, while the connectome's specific wiring holds them at 6.5\% (residual +0.89). The mushroom-body input, feedback and output neurons and the convergence neurons show the same suppression (residuals +0.79, +0.62, +0.44 and +0.62), whereas the descending and premotor classes sit at the block-null level (residuals near zero), their confinement carried by cell-class architecture alone. This localized residual is the part of the routing that no coarse network statistic reproduces, concentrated on the mushroom body, the learning center. The circuit that the finer wiring most specifically keeps out of the sparse drive is the same one that concentrates the dominant driving modes (Fig. 4): two independent signatures of specific wiring in the mushroom body. A lesion control places the global routing confinement upstream of the mushroom body and independent of it (Supplementary Section S6.3).

Two qualifications bound this result. The higher-order statistics show that structure beyond degree and weight is present, and a targeted lesioning of four candidate structural features (reciprocity, modularity, clustering and afferent routing; Supplementary Section S5, structural-feature lesion analysis; the carrier titration and cross-effect controls in Supplementary Tables S7 and S8) finds the confinement distributed across that structure rather than carried by any single feature, so the effect is architectural rather than motif-attributable. This is a property of the frozen operator, not a measurement that sensory input activates a fixed fraction of the living larval brain. Within those limits it is the paper's clearest wiring-specific signature: with each neuron's in-degree, out-degree and outgoing weight matched, the placement of synapses routes sparse input into a tight olfactory shell. The result matters because a frozen operator can separate localization that coarse connection statistics already produce from localization that requires the exact synapse-by-synapse placement. That olfactory circuitry is anatomically localized is itself expected. Restoring each node's in-strength to a rewired graph, the one quantity the matched ensemble leaves free, recovers about 37\% of the confinement magnitude but not its geometry: the active set overlaps the connectome's only partially (Jaccard 0.35), with a different class profile (correlation 0.59) and far weaker concentration (Gini 0.685 against 0.850). These two controls bear on different things. The in-strength-restored graph recovers part of the confinement's magnitude, and the block-preserving null shows cell-class architecture recovers another part. Neither reproduces the concentrated olfactory routing, which, with the residual magnitude, is set by the exact placement of synapses, most of it in the learning center that holds the drive to a fifth of the brain. Figure 6 summarizes the full picture across metrics: the gross operator signature is about 96 to 98\% reproduced by the degree-and-weight statistics, while the sparse-input confinement separates into a cell-class component and a finer within-class component.

\begin{figure}[tbp]
\centering
\includegraphics[width=\linewidth]{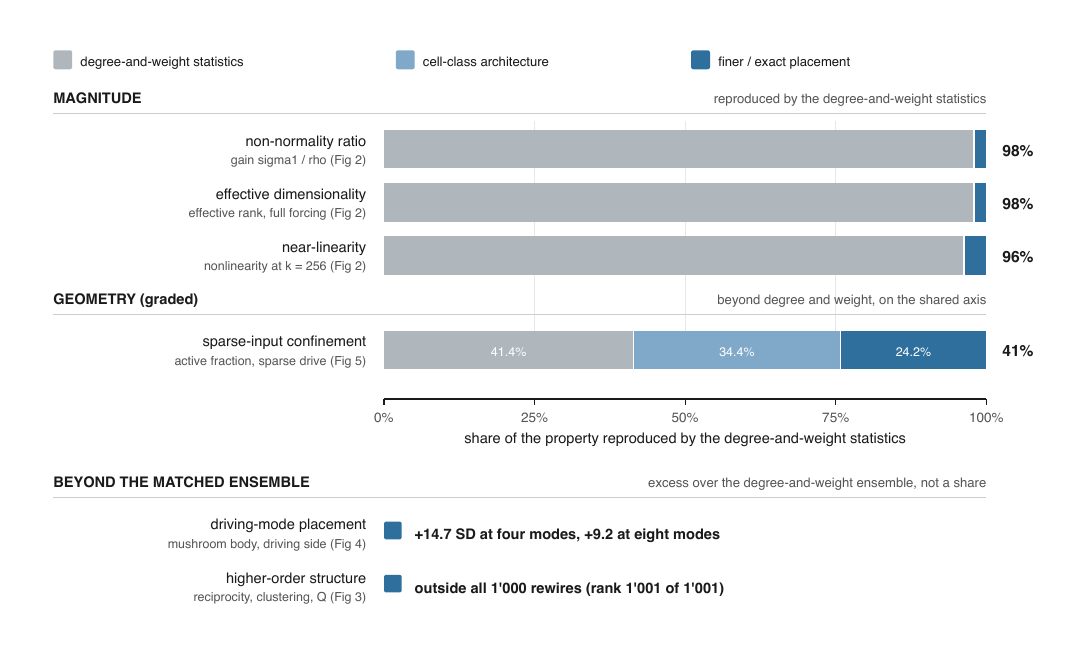}
\caption{\textbf{Degree and weight reproduce the gross operator signature; the exact wiring sets higher-order structure, input routing and mushroom-body modes.} The magnitude properties, operator gain, effective dimensionality at full forcing and near-linearity, are plotted as the share reproduced by the degree-and-weight statistics (grey) against the residual beyond them, and are about 96 to 98\% reproduced. The geometry properties lie beyond the matched ensemble rather than on this shared axis and are shown by their excess over it: the mushroom-body driving-mode concentration exceeds the ensemble by 14.7 and 9.2 standard deviations at four and eight modes, and the higher-order structure ranks outside all 1'000 rewires. Sparse-input confinement is the graded case, with the degree-and-weight statistics reproducing 41\% of it, the cell-class architecture a further 34\%, and the finer within-class placement the remaining 24\%. The cell-class cut and the node-level in-strength control (Supplementary Section S6.1) are complementary diagnostics, not one additive partition; the routing stays placement-specific.}
\label{fig:6}
\end{figure}

\subsection*{2.5 The calyx reconstructs sparse odor coding, but its expansion is generic rather than wiring-specific}

The calyx is the mushroom body's input zone, where a small number of olfactory projection neurons fan out onto a much larger population of Kenyon cells. A classical account holds that this expansion, combined with a firing threshold and feedback inhibition, sparsifies and decorrelates odor representations \citep{turner2008olfactory,honegger2011sparse}. We therefore asked two questions: does the frozen operator reconstruct this sparse-coding function at all, and if it does, is the expansion a product of the exact calyx wiring or a generic consequence of any sparse fan-out with a threshold?

We found that the operator reconstructs the function, in the expected direction. Under a local Kenyon-cell threshold, modeled Kenyon-cell representations are less correlated across odors than their projection-neuron inputs (correlation 0.675 against 0.831), and the linear calyx readout, which on its own compresses dimension (a ratio of 0.66 at eight input modes), inverts to expansion once the threshold is applied. This decorrelation is produced by the local Kenyon-cell threshold alone, with the APL, the circuit's sole GABAergic feedback interneuron and present as the bilateral pair \citep{masuda2014apl,mancini2023apl}, held at zero gain throughout. The reconstruction matches the textbook role of the calyx (Supplementary Table S6, calyx mechanism).

A stronger claim, namely that the exact calyx wiring is what produces this expansion, does not hold up once we average across seeds instead of taking the best one. The connectome's best seed expanded more than all 30 degree-preserving rewires (a ratio of 1.535 against the ensemble maximum of 1.522), which had looked like a wiring-specific effect. Across its three seeds, however, the connectome's expansion ratios were 1.535, 1.318 and 1.406, a mean of 1.420, against a rewire ensemble mean of 1.367 $\pm$ 0.080 (range 1.198 to 1.522). The connectome mean sits only 0.66 standard deviations above the rewire mean, and the two distributions almost entirely overlap. The expansion is therefore generic to a sparse fan-out with a threshold, which is the classical expectation \citep{litwinkumar2017optimal}, rather than a property of the specific calyx wiring. Based on these results, we then retracted the wiring-specific reading and kept the reconstruction.

The calyx thus reconstructs its textbook role in the model, sparse and decorrelated odor coding, while that role follows from the fan-out statistics rather than from the exact wiring. This is the second place where the same rigor applies: a reconstruction the matched ensemble also produces is retained, and the wiring-specific overlay is retracted, as it was for the lateral horn. The model recovers the calyx's sparse-coding function, and that function would arise in any comparably wired fan-out, so the same control hierarchy that keeps the reconstruction is the one that removes the wiring-specific overlay.

\section*{3. Discussion}

Run as a frozen operator with no parameter fitting, the larval connectome separates into two kinds of dynamical property. Its gross response is largely fixed by coarse connection statistics: how strongly it amplifies, how many dimensions it preserves, how nonlinear it is. Where input travels and which neurons dominate the leading modes are fixed instead by the exact wiring. Two candidate localizations did not survive the controls and were retracted. We take the separation first, then the two roles the wiring assigns, then the controls that drew the line.

\subsection*{3.1 The gross response is statistical; routing and the dominant modes are wiring-specific}

The finding that the gross response is statistical is the less surprising half. Almost any wiring with the same degrees and weights reproduces the operator's gain, its dimensionality and its near-linearity, so these belong to the connectome's coarse description and not to its exact circuit. Theory anticipates this, since the dimension of collective activity in a nonlinear recurrent network is set by quantities the matched ensemble preserves \citep{clark2025connectivity}. The more surprising half is that the exact wiring still fixes two properties the statistics leave open, the routing of sparse input and the identity of the leading modes. The confinement of sparse input is the one property that sits between the two halves, recovered partly by cell-class architecture and the rest only by the finer placement of synapses (Fig. 6).

This separation has a counterpart in connectome-constrained models of neural activity. There a connectome alone often fails to fix a network's dynamics, because many single-neuron parameter settings are consistent with the same wiring, and only some of those settings shape the dynamics strongly while others barely change it \citep{beiran2025prediction}. Our operator fits nothing and has no single-neuron parameters, so the counterpart is structural: the matched ensemble is the family of wirings consistent with the connectome's coarse description, and the separation says which dynamical signatures that whole family shares and which the exact wiring alone produces. The gross response is what the family shares, robust to the exact placement in the way that often marks biological systems. The routing and the driving-mode structure are what the placement alone decides.

We use the words regime and geometry, informally, for these two sides: regime for the gross response the statistics fix, geometry for the routing and mode placement the wiring fixes. The terms are shorthand for the metrics defined in Results, and carry no claim beyond them.

\subsection*{3.2 The two wiring-specific roles, as circuit and as operator}

Both wiring-specific results carry a structural and a functional reading at once. As operator structure, the exact wiring fixes a low-dimensional route for sparse input and lifts one subspace into a disproportionate share of the leading driving modes, neither of which the second-order statistics determine. As biology, that same wiring gives the olfactory pathway its narrow channel and the learning center its outsized role in the dominant modes, roles that connection counts alone would not predict. The two signatures land on the same circuit: the finer wiring most specifically keeps the mushroom body out of the sparse drive, and that same circuit concentrates the leading driving modes; a lesion confirms that its grip on those modes runs through its incoming recurrent embedding and not its outgoing weight alone, while placing the global routing confinement upstream of and independent of the mushroom body (Fig. 4, Fig. 5; Supplementary Section S6.3).

Independent structural analyses of this connectome point the same way. A community analysis finds a rich club of hub interneurons concentrated in a few midbrain communities, one aligned with the mushroom body's Kenyon and output cells \citep{betzel2024hierarchical}, and the larval structural core is itself built from mushroom-body neurons \citep{yadav2025rewiring}. Our driving-mode concentration and our olfactory routing are the activity-level counterparts of those structural findings, reached by a different instrument.

\subsection*{3.3 The control battery is a transferable method, and its retractions show its discriminating power}

Drawing the line between the two kinds of property took a battery of nulls, and that battery is both the method we propose and the rigor that held the strongest claims in check. An unstructured operator shows only that the connectome is non-random. A degree-and-weight-matched ensemble asks whether a property needs the exact placement of synapses or only their coarse statistics \citep{maslov2002specificity}. A claim that a named circuit carries a property needs one more comparison, against arbitrary neuron sets of the same size, because in a tight ensemble extreme ranks are common even where nothing is special.

We propose here that the discriminating power of these controls is itself a result, because each removed a candidate that a coarser test had passed. The lateral-horn concentration on the driven side was flagged against the matched ensemble and then removed once arbitrary sets of the same size proved extreme as often. The calyx expansion beat every rewire on its single best seed and fell back inside the ensemble once all three seeds were counted. These retractions give the surviving claims their weight: a property is called wiring-specific only after the matched ensemble and the stricter size-matched and seed controls have failed to reproduce it, so the necessity of these controls is shown on real candidates and not asserted.

\subsection*{3.4 Scope and outlook}

These are statements about a frozen, rate-based, norm-rescaled operator. It omits action potentials and single-neuron time constants, graded and short-term synaptic dynamics, receptor kinetics, neuromodulatory and behavioral state, and gap junctions, and it fits no biophysical parameter, so it does not report the dynamics of the living animal. It reports which features of the reconstructed wiring control which measured properties of the operator, and that is the statement the controls support. We make no spatial claim, since the operator has no geometric term and straight-line distance poorly predicts single-neuron connectivity in this connectome \citep{betzel2024hierarchical}, so the placement that produces the confinement may itself be spatially organized in ways our nulls neither model nor need to separate.

For the larval \textit{Drosophila} connectome the result is definite. Its global operator behavior is mostly statistical, its dominant-mode leverage is diffuse apart from the mushroom body's over-concentration on the driving side, and the clearest mark of its specific wiring is the routing of sparse input into a compact olfactory shell, beyond cell-class architecture. Whether the same split holds in other connectomes is open, and the same matched-null approach can settle it, because the ensembles travel with the graph and not with the species, and the same logic can audit an engineered recurrent architecture as readily as a brain. What we offer is a portable measurement: a clean partition of a connectome's dynamics into the part its wiring alone determines and the part its statistics already fix. The larval brain is where we first read it off.

\section*{4. Methods}

\subsection*{4.1 Known confounds and scope}

Several limitations bound every result below and are collected here. Synapse signs are not used: the recurrent matrix is built from unsigned synapse-count magnitudes, so excitation and inhibition are not distinguished (the routing result is shown to survive an excitation-inhibition sign cone). The matched ensembles preserve out-strength but not each neuron's in-strength, so the sparse-input confinement is decomposed against a row-strength-restored control, which recovers part of its magnitude while its geometry remains placement-specific. Every operator is rescaled to a common spectral radius, so absolute gain is normalized away and the comparisons are of geometry at a fixed scale (the routing result persists across a grid of spectral radius, leak and input amplitude). Activity is thresholded at a fixed fraction of the driven-set variance, and the confinement holds across that threshold. The leading-mode concentration is a spectral, adjoint-side property, not a demonstration of causal control. All results are for a single connectome, the larval \textit{Drosophila}; whether they generalize is an open question. The model is a frozen operator, not a measurement of activity in the living animal.

\subsection*{4.2 Connectome dataset and recurrent core}

We used the complete larval \textit{Drosophila} connectome \citep{winding2023connectome}, taking the directed, weighted edge list in which each edge weight is the reconstructed synapse count between two neurons. The graph has 3'013 neurons and 111'243 directed synaptic edges; weights are integer synapse counts and are used as magnitudes only, without sign. Data integrity was checked by SHA-256 against the archived manifest before use.

The recurrent substrate is the largest strongly connected component of this graph, computed independently with two libraries that agreed neuron for neuron: a core of 2'825 neurons joined by 109'438 internal directed edges. Reported neuron and connection counts for this connectome vary with the processing version: the core here is 2'825 neurons, against the 2'810-node strongly connected component in \citep{seguin2025connectome} and the 2'952 cells in \citep{betzel2024hierarchical}. The remaining 188 neurons form a thin feedforward periphery that partitions cleanly into 80 afferent ports (they send into the core and receive nothing back, through 524 couplings) and 97 efferent ports (they receive from the core and send nothing in, through 1'161 couplings), with the remaining 11 peripheral neurons connecting only within the periphery and no neuron serving both an afferent and an efferent role. The counts close on both sides: 2'825 + 80 + 97 + 11 = 3'013 neurons, and 109'438 core-internal, 524 afferent-injection, 1'161 efferent-readout and 120 periphery-internal couplings sum to the 111'243 directed connections. Within the core, 536 self-loops (autapses, weights 1 to 14) are retained on the diagonal, and 235 core neurons receive afferent input. The olfactory receptor neurons lie inside the core rather than on the afferent shell, so a port-only drive enters the olfactory pathway downstream of the receptor layer, a point we return to in the routing assay.

\subsection*{4.3 Frozen rate operator}

The core is run as a frozen recurrent rate operator with no trained or fitted parameters. The core weight matrix W is built from the core-internal edges with the convention W[i, j] equal to the synapse count from source neuron j to target neuron i; W is rescaled so that its spectral radius equals $\rho^\star = 0.99$ (raw spectral radius 296.6, largest singular value 393.1, Frobenius norm 1'825.2):

\begin{equation}
W \in \mathbb{R}^{N \times N},\quad W_{ij} = c(j\!\to\!i),\quad \tilde W = \frac{\rho^\star}{\rho(W)}\,W,\quad \rho(W) = \max_i \lvert \lambda_i(W)\rvert,\quad \rho^\star = 0.99,
\end{equation}

where $N$ = 2'825 and $c(j\!\to\!i)$ is the synapse count from neuron $j$ to neuron $i$. Neural state evolves by the leaky-tanh update

\begin{equation}
\mathbf{x}_n = (1-\alpha)\,\mathbf{x}_{n-1} + \alpha\,\tanh\!\big(\tilde W \mathbf{x}_{n-1} + B\,\mathbf{s}_n\big), \qquad \alpha = 0.9,
\end{equation}

with $\tanh$ applied elementwise and the input matrix $B$ and input stream $\mathbf{s}_n$ defined per assay below. Every measured property is therefore a function of the wiring and the synapse-count weights alone. The single-neuron parameters of a trained network (per-neuron gains and biases) are absent by construction, which is what makes the degree-and-weight-matched ensemble below the appropriate control: it isolates the contribution of exact synaptic placement from that of the degree and weight statistics.

\subsection*{4.4 Global operator metrics}

Three scalar properties summarize the operator. The non-normality ratio $\nu$ is the ratio of the largest singular value to the spectral radius, computed by sparse singular-value and eigenvalue solvers; for a normal operator this ratio is 1, and its excess measures non-normal transient amplification.

\begin{equation}
\nu = \frac{\sigma_1(\tilde W)}{\rho(\tilde W)},
\end{equation}

with $\sigma_1$ the largest singular value and $\rho$ the spectral radius (the largest eigenvalue magnitude; $\nu = 1$ for a normal operator). The effective dimensionality is the participation ratio of the centered state covariance, evaluated when the operator is driven by an input of fixed rank k; we report it at k = 8 and at full rank.

\begin{equation}
\mathrm{PR} = \frac{\big(\textstyle\sum_i \lambda_i\big)^2}{\sum_i \lambda_i^2},
\end{equation}

with $\lambda_i$ the eigenvalues of the centered state covariance $C = \tfrac{1}{T}\,X_c^{\top} X_c$ ($X_c$ the centered $T \times N$ state matrix), at input rank $k$. This participation ratio is the dimension-of-activity statistic whose dependence on the coupling variance and the effective rank of the connectivity is derived for nonlinear recurrent networks by \citep{clark2025connectivity}. Near-linearity is the fraction of state variance, in the operator's own top-256 principal subspace, that is not linearly reconstructable from the input stream and its recent history; a small value means the operator acts almost linearly on its inputs.

\begin{equation}
f_{\mathrm{nl}} = 1 - \frac{\lVert Q^{\top} X_c\rVert_F^2}{\lVert X_c\rVert_F^2},
\end{equation}

where $Q$ is an orthonormal basis of the centered input regressors $\{\mathbf{s}_n,\mathbf{s}_{n-1},\dots,\mathbf{s}_{n-d}\}$ (history depth $d = 0$, the current input only) and $X_c$ is restricted to its top-256 principal subspace; $f_{\mathrm{nl}}$ is reported at $k = 256$. The dimensionality and near-linearity assays use a fixed drive regime: a round-robin rank-k input matrix that holds per-node input power constant across k (amplitude 0.3), independent white-noise streams, a washout followed by a long measurement window. Henrici's departure from normality and a covariance-eigenvalue path are computed as independent cross-checks. The relation between the largest singular value and the row and column mass of $W$ that bounds $\sigma_1$ follows standard matrix-norm inequalities \citep{hornjohnson2012matrix}.

\begin{equation}
\sigma_1(W) \le \sqrt{\,\lVert W\rVert_1\,\lVert W\rVert_\infty\,} = \sqrt{\Big(\max_j \textstyle\sum_i \lvert W_{ij}\rvert\Big)\Big(\max_i \textstyle\sum_j \lvert W_{ij}\rvert\Big)},
\end{equation}

relating the operator 2-norm to the maximum column mass and row mass. Because the degree-and-weight ensemble preserves out-strength, it preserves the column-mass factor of this bound exactly, so the reproduction of the non-normality ratio is in part definitional rather than discovered. Henrici's departure from normality, $\Delta(W) = \big(\lVert W\rVert_F^2 - \sum_i \lvert \lambda_i\rvert^2\big)^{1/2}$, is computed as a corroborating cross-check.

\subsection*{4.5 Null models}

Three nulls, ordered from weakest to strongest, support the inferences.

The unstructured Gaussian control is a random directed matrix with the same node count and the same number of edges as the core, with independent standard-normal weights placed uniformly at random and rescaled to spectral radius $\rho^\star$. It matches the connectome only in size, edge density and spectral radius, and it serves to detect gross departures from randomness.

The degree-and-weight-matched ensemble is the central control. Each instance is generated from the connectome by directed double-edge swaps (target ten times the edge count), which preserve every neuron's in-degree and out-degree, each neuron's out-strength (its total outgoing weight, because weights travel with their source), the entire multiset of synapse-count weights and the diagonal self-loops exactly, while scrambling the placement of edges. Each neuron's in-strength (its total incoming weight) is therefore not separately preserved, nor is any higher-order structure such as cell-class blocks, hemisphere or reciprocity.

\begin{equation}
d^{\mathrm{in}}_i = \sum_j A_{ij},\qquad d^{\mathrm{out}}_j = \sum_i A_{ij},\qquad \{\,W_{ij} : W_{ij} \neq 0\,\}\ \text{(multiset)},
\end{equation}

all preserved exactly by each swap, with $A$ the binary adjacency of $W$. Because a rank-based test resolves the connectome's position in the null only as finely as the ensemble is large, we use 1'000 instances (seeds 2000 to 2999); the first 50 are cached and reproduce the values of the earlier 50-instance runs, and the move to 1'000 follows the network-ensemble resolution standard for this kind of comparison. A property the ensemble reproduces is set by the degree and weight statistics; a property on which the connectome falls outside the ensemble requires the exact placement of synapses \citep{maslov2002specificity}. Because this null preserves only degree and the weight multiset and demonstrably destroys the connectome's community structure (the higher-order statistics above), it does not retain the annotation that community-preserving or block-preserving nulls carry \citep{betzel2024hierarchical}.

The random size-matched set control supports the identity-resolved leverage analysis only. For each functional set it draws 1'000 random sets of neurons of the same size (seed 19019) and asks how often an arbitrary set of that size is as extreme as the named circuit.

So that the confinement result is tested against cell-class architecture and not merely against degree, for the sparse-drive confinement we add a block-preserving null, in the tradition of community-preserving graph nulls, that holds the cell-class architecture fixed. It is the degree-and-weight rewire restricted so that both swapped edges share the cell class of their source and the cell class of their target, so swaps stay inside each cell-class block; it therefore preserves exactly, against the connectome, every neuron's in-degree and out-degree, each neuron's out-strength, the full weight multiset and the 18-by-18 class-to-class edge-count matrix, while scrambling placement within blocks. We use 1'000 instances (seeds 2000 to 2999, paired with the degree-and-weight ensemble) over an 18-class partition of the core. The within-block mixing is near-complete (edge displacement 0.835; frozen-edge fraction below 0.0001), so the null is not conservative by construction. A property the block null reproduces is set by cell-class architecture; a property on which the connectome falls outside it requires finer, within-class placement.

\subsection*{4.6 Higher-order structural statistics}

To characterize structure beyond degree and weight, we computed five graph statistics on each instance with a fixed tool: directed binary reciprocity (the fraction of directed edges whose reverse is also present, self-loops removed), global clustering (transitivity) and average local clustering on the undirected projection, Louvain modularity (resolution 1, seed 50503) and the total triangle count.

\begin{equation}
r = \frac{\sum_{i\neq j} A_{ij} A_{ji}}{\sum_{i\neq j} A_{ij}},\qquad C = \frac{3\,n_{\triangle}}{n_{\wedge}},\qquad \bar C = \frac{1}{N}\sum_i \frac{2 e_i}{k_i(k_i - 1)},
\end{equation}

with $A$ the directed binary adjacency (self-loops removed), $A^{u}$ its undirected projection of $m$ edges and degrees $k_i$, $n_{\triangle}$ the triangle count, $n_{\wedge}$ the connected-triple count, $e_i$ the edges among neuron $i$'s neighbors, and $c_i$ the Louvain community of $i$.

\begin{equation}
Q = \frac{1}{2m}\sum_{ij}\Big(A^{u}_{ij} - \frac{k_i k_j}{2m}\Big)\,\delta(c_i, c_j),
\end{equation}

These are the properties the degree-and-weight ensemble destroys, and they license reading the routing result as a structure-beyond-degree-and-weight effect.

\subsection*{4.7 Module maps}

The later analyses ask whether named circuits carry particular dynamical properties, so each circuit is defined explicitly and its boundary sharpness is stated. Cell identities come from the dataset class labels. The mushroom body is unambiguous: its connectivity-clustering membership and its cell-class labels coincide at 231 neurons, comprising 153 Kenyon cells, 48 output neurons, 28 modulatory input neurons and the two APL neurons, the bilateral pair \citep{eichler2017mbconnectome}. The lateral horn and the central complex are bounded less sharply. For each we report a clustering-defined set (the core-restricted cell-class set that matches the connectivity-clustering atlas) and a broader expert-annotation set: the lateral horn at 201 (clustering) versus 445 (annotation), intersection 164 \citep{berck2016olfactory,eschbach2020recurrent}, and the central complex at 77 versus 34, overlap 10 \citep{lungu2025larvalcx}.

\begin{equation}
S^{\mathrm{clust}}_{\mathrm{LH}} = \{\,i : \mathrm{class}(i) = \text{LHN}\,\},\qquad \lvert S^{\mathrm{clust}} \cap S^{\mathrm{annot}}\rvert = 164\ (\text{LH}),\quad 10\ (\text{CX}).
\end{equation}

Every later claim states which boundary it uses. The community-ordered connectivity matrix that displays the modular block structure uses the Louvain partition above.

\subsection*{4.8 Input-drive assays}

All driven measurements use the leaky-tanh update above with white-noise input streams (independent standard-normal sequences), a washout of 1'000 steps and a measurement window of 20'000 steps. Two input constructions are used. For the dimensionality and near-linearity metrics, B is the round-robin rank-k partition matrix that holds per-node input power constant as k varies, so only input rank changes. For the routing assay, B is the afferent injection matrix built from the real 524 afferent-to-core couplings, scaled by the synapse counts, with one independent stream per afferent port; this drives the 235 core neurons that receive afferent input.

The input streams are independent and white, $\mathbf{s}_n \sim \mathcal{N}(0, I_k)$. The two input matrices are

\begin{equation}
B^{\mathrm{rank}}_{i,\,(i \bmod k)} = a \ \ (\text{otherwise } 0),\qquad B^{\mathrm{aff}}_{ip} = \mathrm{amp}\cdot c(p\!\to\!i)\ \ (p\ \text{an afferent port}),
\end{equation}

so $B^{\mathrm{rank}}$ has rank $k$ and gives every node input variance $a^2$ for all $k$ (amplitude $a = 0.3$), while $B^{\mathrm{aff}}$ injects through the real afferent couplings.

\subsection*{4.9 Sparse port-drive confinement}

The routing assay drives the core through its 80 afferent ports at amplitude 0.10 (raw synapse counts times the amplitude), one white-noise stream per port (stream seed 93101), at spectral radius 0.99, leak 0.9, washout 1'000, window 20'000. A neuron is counted active if its temporal state standard deviation exceeds one tenth of the median standard deviation of the directly-driven set. The active fraction is the share of the 2'825 core neurons that meet this threshold.

\begin{equation}
\mathrm{sd}(x_i) = \Big[\tfrac{1}{T}\sum_n (x_{i,n} - \bar x_i)^2\Big]^{1/2},\quad \theta = \tfrac{1}{10}\,\operatorname{median}_{j \in D}\,\mathrm{sd}(x_j),\quad \varphi = \frac{1}{N}\big\lvert\{\,i : \mathrm{sd}(x_i) > \theta\,\}\big\rvert,
\end{equation}

with $D$ the directly-driven set; the connectome gives $\varphi = 0.1986$ (561 of 2'825). We report the connectome active set (561 neurons, 0.1986), the active fraction of each cell class and each module, and the activation binned by undirected hop distance from the driven set. The identical ports and streams drive each of the 1'000 degree-and-weight-matched instances and the Gaussian control, so the three differ only in wiring.

\subsection*{4.10 Identity-resolved mode leverage}

To ask which neurons shape the operator's dominant modes, we computed a per-neuron leverage on three subspaces. For a subspace given by an orthonormal real basis U (columns), the leverage of neuron i is the squared row norm of its entries, the statistical leverage that measures how strongly that neuron loads onto the subspace. The three subspaces are the dominant m-dimensional invariant subspace of W (the right-eigenvector or driven subspace), the same for W transpose (the left-eigenvector or driving subspace), and the top-m left singular subspace, with m taken at 1, 2, 4, 8 and 16 and complex conjugate pairs completed into real planes. For each of seven pre-specified, annotation-defined neuron sets, the leverage energy on the set is the leverage summed over its neurons divided by the total.

\begin{equation}
\ell_i = \sum_{j=1}^{m} U_{ij}^2 = (U U^{\top})_{ii},\qquad \sum_i \ell_i = m,\qquad E_S = \frac{\sum_{i \in S} \ell_i}{\sum_i \ell_i},
\end{equation}

with $U \in \mathbb{R}^{N \times m}$ an orthonormal real basis of the subspace and $S$ a neuron set. The seven sets are the mushroom body (231), the lateral horn (201), the central complex (77), the convergence neurons (204, of which 184 lie outside the three mapped modules), the descending output neurons (349), the mushroom body's recurrent output network (its 76 output and modulatory neurons) and the 561-neuron sparse-drive active set. Each energy is computed on the connectome and on the 1'000-instance ensemble, giving a rank and a z-score per set, subspace and m. Because the driving side uses the left eigenvectors of a strongly non-normal operator, the five certified entries were stress-tested: a dense eigensolver reproduces the sparse leverage to absolute difference 0.0000, the dominant eigenvalues are well-conditioned (top-eight condition numbers from 1.1 to 8.2, all below 10), and re-running the sparse solver from five independent start vectors leaves every value unchanged to five decimals.

\subsection*{4.11 Random-set and singular guards, and the lateral-horn retraction}

Because a named set can rank extreme against the ensemble merely by virtue of its size or its overlap with a generic singular direction, a circuit is read as carrying a property only after two guards. The random size-matched guard, a size-matched permutation control, draws 1'000 random sets of the circuit's size (seed 19019) and computes, per subspace and m, the 95th percentile of the absolute z-score across random sets and the fraction of random sets that are rank-extreme; an entry passes only if it is rank-extreme against the ensemble, its absolute z exceeds the random 95th percentile, and fewer than 20\% of random sets are themselves extreme. The singular guard flags an entry as leakage if it is extreme while the same set's singular entry at that m is extreme in the same direction, so that an apparent eigen-subspace effect that is really a generic singular-subspace effect is not counted. An entry is wiring-specific only if it passes the random guard and is not flagged by the singular guard.

\begin{equation}
\text{wiring-specific} \iff (\text{rank extreme}) \,\wedge\, \big(\lvert z\rvert > z^{\mathrm{rand}}_{95}\big) \,\wedge\, \big(p^{\mathrm{rand}}_{\mathrm{ext}} < 0.20\big) \,\wedge\, (\text{no singular leakage}),
\end{equation}

where $z^{\mathrm{rand}}_{95}$ is the 95th percentile of $\lvert z\rvert$ over the 1'000 size-matched random sets and $p^{\mathrm{rand}}_{\mathrm{ext}}$ the fraction of those sets ranked extreme. Five entries survive: the mushroom body over-concentrates the driving (left) modes at m = 4 and m = 8 (the two strongest, with 4.7\% and 7.1\% of random sets extreme), and three further entries (convergence neurons at m = 8 and m = 16, lateral horn at m = 16) lie on the driving side, two of them at the 20\% random-extreme boundary. We therefore lean on the three entries comfortably inside the boundary. A lateral-horn localization on the driven (right) side passed the singular guard but failed the random guard at the mode counts where random sets of size 201 are themselves frequently extreme; we retract that localization and report the driven-side lateral-horn result as not wiring-specific. Because scanning the full 105-entry panel for the most extreme entry inflates the chance of a false positive, we add, in the multiple-comparisons tradition, a family-wise control across the full 105-entry panel (seven sets, three subspaces, five mode counts), a Westfall-Young single-step max-statistic permutation: for each rewire instance we compute a leave-one-out z for every entry and take the family maximum of its absolute value, and the family-wise-adjusted p-value of a connectome entry is one plus the number of rewire instances whose family maximum reaches the connectome's absolute z, divided by one thousand and one. The mushroom-body driving-side entries survive at adjusted p = 0.002 (m = 4) and 0.003 (m = 8); the five flagged entries all fall below 0.05, while the two highest-mode-count entries stay marginal under the size-matched guard.

\subsection*{4.12 Calyx mechanism assays}

The calyx assays test whether the model reconstructs sparse-coding function and whether any reconstruction is specific to the exact calyx wiring. On the olfactory subset (receptor, projection and Kenyon-cell populations), pattern separation is measured as the mean pairwise correlation of Kenyon-cell representations across odors compared with that of their projection-neuron inputs (0.675 versus 0.831), and the calyx readout dimensionality is measured under two readouts: a linear readout, which compresses (a ratio of 0.66 at eight input modes), and a Kenyon-cell threshold with feedback inhibition, which inverts this to expansion \citep{turner2008olfactory,honegger2011sparse}.

\begin{equation}
\bar c = \frac{2}{P(P-1)}\sum_{a<b}\frac{\langle \tilde{\mathbf r}_a, \tilde{\mathbf r}_b\rangle}{\lVert \tilde{\mathbf r}_a\rVert\,\lVert \tilde{\mathbf r}_b\rVert},\qquad \mathrm{XR} = \frac{\mathrm{PR}(\text{KC})}{\mathrm{PR}(\text{PN})},\qquad x^{\mathrm{KC}}_n = \tanh\!\big(\mathrm{relu}(z^{\mathrm{KC}}_n - \theta)\big),
\end{equation}

with $\mathbf r_a$ the population response to odor $a$ ($\tilde{\ }$ centered), $\mathrm{PR}$ the participation ratio above and $z^{\mathrm{KC}}$ the Kenyon-cell pre-activation; the linear readout gives $\mathrm{XR} \approx 0.66$ and the thresholded activation inverts it to $\mathrm{XR} > 1$. To test wiring specificity we rewired only the projection-to-Kenyon-cell edges by degree-preserving double-edge swaps (a 30-instance ensemble). The connectome's expansion ratio across its three seeds (1.535, 1.318, 1.406; mean 1.420) lies 0.66 standard deviations above the rewire ensemble mean (1.367 plus or minus 0.080), with almost fully overlapping ranges, so the expansion is generic to a sparse thresholded fan-out \citep{litwinkumar2017optimal}. We retain the reconstruction and retract the wiring-specific reading of the expansion. The APL appears in the dataset as two reconstructed skeletons, the bilateral pair with one in each hemisphere, which is the same cell type classically described as the single GABAergic feedback interneuron of each mushroom body \citep{masuda2014apl,mancini2023apl}. These calyx metrics are computed on the pooled bilateral module: the 153 Kenyon cells of both calyces are treated as one population, the APL pair is scaled by a single shared inhibitory gain, and one coincidence threshold is applied across the population (the correlation is taken at zero APL gain, where the threshold alone decorrelates). Recomputed per hemisphere, with each calyx as an independent ipsilateral module under its own threshold and its own APL of the pair while the contralateral APL is ablated (its column set to zero, since the operator is magnitude-only), the reconstruction holds on both sides: Kenyon-cell correlation stays below projection-neuron correlation (left 0.675 against 0.759, right 0.741 against 0.889) and the compressive linear readout inverts to expansion (left 0.669 to 1.243, right 0.683 to 1.455), so the pooled reading does not depend on combining the two calyces.

\subsection*{4.13 Statistics}

Each property is summarized by its rank among the 1'001 graphs (the connectome and its 1'000 matched rewires) and by a z-score, the connectome value minus the ensemble mean over the ensemble standard deviation. Significance of a wiring-specific claim is read against the level-matched null: the Gaussian control for departures from randomness, the degree-and-weight ensemble for exact placement beyond coarse statistics, and the random size-matched sets for circuit-level localization.

\begin{equation}
z = \frac{v_{\mathrm{conn}} - \mu_{\mathrm{ens}}}{\sigma_{\mathrm{ens}}},\qquad \mathrm{rank} = 1 + \big\lvert\{\,e : v_e < v_{\mathrm{conn}}\,\}\big\rvert,
\end{equation}

with $\mu_{\mathrm{ens}}, \sigma_{\mathrm{ens}}$ the ensemble mean and standard deviation, $v_e$ the ensemble values, and the rank taken among the 1'001 graphs. No distributional assumptions are placed on the ensembles; ranks and the empirical null distributions carry the inference.

\subsection*{4.14 Reproducibility, audits and provenance}

The operator is built deterministically with no random number generator; every stochastic step uses a fixed, recorded seed (rewire instances 2000 to 2999, random sets 19019, port streams 93101, Louvain 50503, rank-k sweep streams 70000 plus k). Each reported value reproduces on a fresh process from the recorded seeds, and the headline numbers were recomputed at 1'000-instance resolution. Every principal interpretive claim was checked against its matched null; two claims that did not survive (the driven-side lateral-horn localization and the wiring-specific calyx expansion) were retracted, with the original text preserved alongside the retraction. Data provenance is fixed by SHA-256 against the archived manifest. The derived result files, with an integrity manifest and seeds, are publicly available at https://github.com/Stavros963/fcta, and the full analysis pipeline follows on publication; the connectome itself is publicly available from its original source \citep{winding2023connectome}.

\section*{Data availability}

The primary data are the publicly available larval \textit{Drosophila} connectome and its associated public annotations \citep{winding2023connectome}. Derived analysis results, with an integrity manifest, random seeds and a verification script, are publicly available at https://github.com/Stavros963/fcta.

\section*{Code availability}

A public verification bundle, containing the result files behind every figure, the figures, the random seeds, the environment lock and an integrity manifest, is available at https://github.com/Stavros963/fcta. The full analysis pipeline will be released on publication and is available from the author on reasonable request before then.

\section*{AI-use disclosure}

During the preparation of this manuscript, the author utilized Anthropic Claude to assist with adversarial critique, code scaffolding for manuscript checks, and prose clarity. This technology was used strictly as an assistive aid, and no AI tool is listed as a co-author. All manuscript figures are entirely deterministic and data-derived; no generative AI was used to create or alter visual data. The author retains full accountability for the integrity of the work, and all scientific claims, quantitative metrics, analyses, and interpretations were manually cross-verified against the underlying code, raw data, and the analysis records.

\section*{Acknowledgements}

We thank the teams that reconstructed and openly released the larval \textit{Drosophila} connectome and its associated annotation resources \citep{winding2023connectome}, without which this study would not have been possible.

\section*{Competing interests}

The author declares no competing interests.

\bibliographystyle{unsrtnat}
\bibliography{references}

\end{document}